\documentclass[11pt,a4paper]{article}
\usepackage[utf8]{inputenc}
\usepackage{amsmath, bm}
\usepackage{amsfonts}
\usepackage{amssymb}
\usepackage{tikz}
\usetikzlibrary{matrix,chains,positioning,decorations.pathreplacing,arrows}
\usepackage{graphicx, color}
\usepackage{breqn}
\usepackage{booktabs,adjustbox}
\usepackage{booktabs}
\usepackage{float}
\usepackage{epstopdf}
\usepackage{titlesec}
\usepackage{rotating, changepage}
\usepackage{atbegshi,picture,subfigure}
\titleformat*{\subsection}{\normalfont}

\author{Jos\'e Igor Morlanes}
\title{Non-linear Time Series and Artificial Neural Network of Red Hat Volatility.}

\begin{document}

\maketitle

\begin{abstract}
We extend the empirical results published in the paper \emph{Empirical Evidence on Arbitrage by Changing the Stock Exchange} \cite{morlanes2009empirical} by means of machine learning and advanced econometric methodologies based on Smooth Transition Regression models and Artificial Neural Networks.
\end{abstract}

\section{Introduction}
In this paper we examine wether there is a reduction in the Red Hat inc.  stock volatility during the moving from NASDAQ to New York Stock Exchange (NYSE) on December 12, 2006 \cite{morlanes2009empirical}. We model the dynamics of the volatility by means of non-linear autoregressive models and machine learning approach. We mainly focus on three models: the Logistic smooth transition regression model (LSTAR), the self-exciting threshold autoregressive model (SETAR), and the neural network non-linear autoregressive model (NNET). 

NASDAQ and NYSE are markets in which trade take place under very different conditions. It seems natural to assume that Red Hat Inc. stock dynamics suffers a change when markets are switched. We naturally allow the stock price to consist of two different regimes or states of the world and allow the dynamics to be different in the two regimes. One before the switch of the markets and other after the switch of markets. Classical linear models seem not to capture the complexity of this change.  Non-linear models may be more appropriate \cite{franses2000non}. 

A popular set of models applied in different regimes are autoregressive (AR) models such as SETAR and LSTAR models. These models are extentions of the linear AR models. They are easily estimated and intepreted using regression methods.

We explore machine learning approach  as an alternative semiparametric method. The use of NNET has become very popular in the last two decades. This is due to its capacity of learning the ''hidden'' relationships in the data without the necessity of supposing a particular parametric model. We confine ourselfves to the applications of Artificial Neural Networks (ANN) and do not consider other types of machine learning approches such as support vector machines and other kernel based learning methods.

\section{Data set}

The data set includes 500 observations of daily closing prices of Red Hat financial assets. These daily prices are sourced from the Federal Reserve Bank of St. Louis Economic Data (FRED). 

Unit root test, based on the non-linear Perron test, indicates that the time series is non-stationary. We therefore choose to work with the first difference of the logarithmic price. To perform the non-linear Perron test, we first consider a one-time structural break at $T_B="December 12, 2006"$ with $1<T_B<T$. The null hypothesis consists of a unit root with possible non-zero drift which permits a structural change in the level and the growth rate of the price series
\begin{equation}
p_t = \mu_1+p_{t-1}+dD(TB)_t+(\mu_2-\mu_1)DU_t+e_t
\end{equation}
where
\begin{equation*}
D(TB) = \begin{cases}
             1  & \text{if } t=T_B+1 \\
             0 & \mathrm{otherwise}
       \end{cases} \quad \mathrm{and}\quad
DU_t = \begin{cases}
             1 & \text{if } t > T_B \\
             0 & \mathrm{otherwise}
       \end{cases}
\end{equation*}
versus the alternative hypothesis of a trend-stationary model which allows one change in the intercept and one change in the slope of the trend function 
\begin{equation}
p_t = \mu_1+\beta_1 t+(\mu_2-\mu_1)DU_t+(\beta_2-\beta_1)DT_t+e_t
\end{equation}
where
\begin{equation*}
DT_t = \begin{cases}
             t-T_B & \text{if } t>T_B \\
             0 & \mathrm{otherwise}.
       \end{cases}
\end{equation*}
To motivate the particular choice of the hypothesis test, we illustrate in Figure \ref{RH_series}(a), the trend of the Red Hat Inc.  

After detrending the price series, we perform a Phillips-Perron test on the residulas $e_t$. We do not reject the unit root hypothesis with Z-statistic -2.8112 and p-value 0.235. We therefore use the logarithm of the first differences of the price series.

\begin{equation*}
y_t=\log p_{t+1}-\log p_t
\end{equation*}

\begin{figure}[H]
\begin{center}
\subfigure[t][Structural break.]{
\resizebox*{5.8cm}{!}{\includegraphics[scale=0.15]{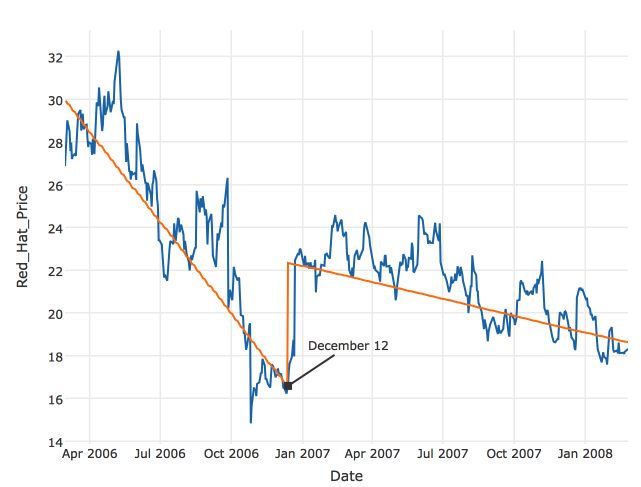}}}\hspace{-5pt}
\subfigure[t][Logarithmic returns.]{
\resizebox*{5.8cm}{!}{\includegraphics[scale=0.15]{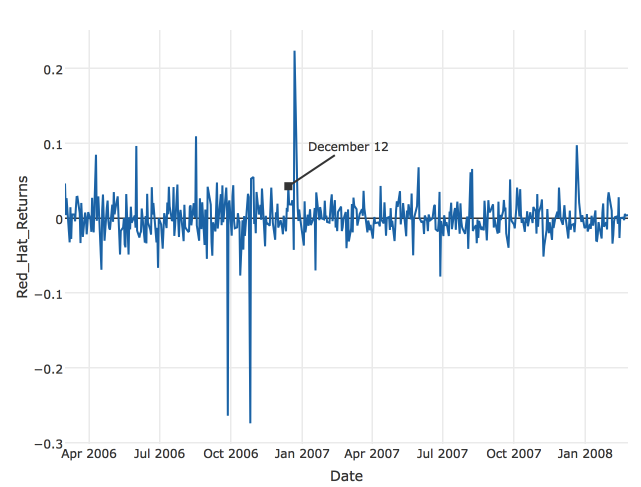}}}
\caption{Time series plots of Red Hat Inc. stock price and returns. (a) The trend of the price series shows a jump  and a change of growth rate at December 12, 2006. (b) Illustrates  a possible reduction in the fluctuations of the returns after the switch of markets on December 12, 2006.}
\label{RH_series}
\end{center}
\end{figure}

We construct the realized volatility time series from the log returns with a window of 60 days. The time series is smooth and tractable for modelling. The series has a clear two regime with a definite structural break at the time of market switch (see Figure 1). We consider other window alternatives such as monthly or quarterly - 30 and 90 days respectively-. Although they have clear economical meaning, they do not produce volatility trajectories which are easily to model. The 30 days window produces a too wild fluctuation series and the 90 days window produces a too short series for meaningful statistics.

\begin{figure}[H]
\begin{center}
\includegraphics[scale=0.35]{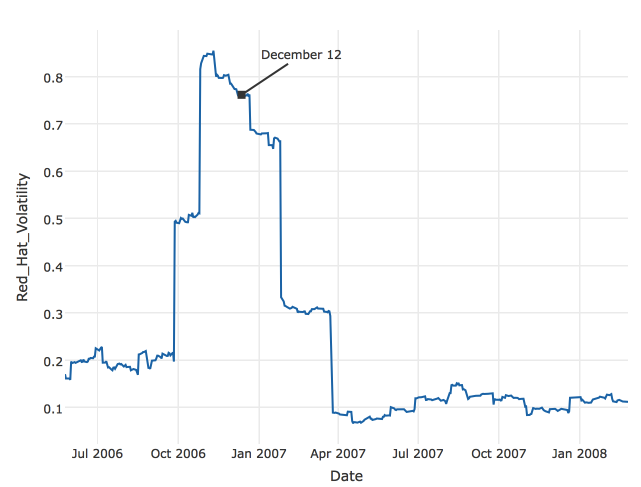}
\caption{Realized volatility with a 60 days window.}
\end{center}
\label{RH_Volatility}
\end{figure}

\section{Econometric Methods}

We use a non-linear autoregressive time series model in the analysis. Consider a general  time series autoregressive model that is generated by
\begin{equation*}
X_t=f(y_t,y_{t-1},\ldots,y_{t-p}; \theta)+\varepsilon_t
\end{equation*}
with $f$ a generic fuction from $R^p$ to $R$. The vector $\theta$ indicates a generic vector of parameters governing the shape of $f$, which are estimated on the basis of an observed time series.

A classical autoregressive model (AR) model is specified by
\begin{equation*}
X_t=\phi+\phi_0 X_{t-1}+\ldots \phi_p X_{t-p}+\varepsilon_t. 
\end{equation*}
A Self-Exciting Threshold Autoregressive Model (SETAR) can be written as:
\begin{equation*}
X_t = \begin{cases}
          \phi+\phi_0 X_{t-1}+\ldots \phi_p X_{t-p}+\varepsilon_t.   & \text{if } X_{t-1}> c \\
         \beta+\beta_0 X_{t-1}+\ldots \beta_p X_{t-p}+\varepsilon_t & \text{if } X_{t-1}< c.
       \end{cases}
\end{equation*}

A Smooth Transition Autoregressive model (STAR) can be viewed as a generalisation of a SETAR model. This allows to change the autoregressive parameters slowly and can be written as:
\begin{equation}
\label{STARmodel}
X_t=\phi+\phi_0 X_{t-1}+\ldots \phi_p X_{t-p}+G(Z_t;\gamma,c)(\beta+\beta_0 X_{t-1}+\ldots \beta_p X_{t-p})+\varepsilon_{t}.
\end{equation}
If
\begin{equation}
\label{Logistic_function}
G(Z_t;\gamma,c)=\frac{1}{1+e^{-\gamma(Z_t-c)}},\quad \gamma>0,
\end{equation}
the logistic function and $Z_t$ is the threshold variable, the model is called Logistic Smooth Transition model (LSTAR). The parameter $c$ can be interpreted as the thereshold and $\gamma$ determines the speed and smoothness of transition.  
The exponential form of the model (ESTAR) uses equation (\ref{STARmodel}) with 
\begin{equation*}
\label{Exponential_function}
G(Z_t;\gamma,c)=1-e^{-\gamma(Z_t-c)^2},\quad \gamma>0.
\end{equation*}

In the empirical study, we use the approach of Artificial Neural Networks models (ANN). A neural network model with linear input, D hidden units and activation function $g$, can be written as:

\begin{equation*}
X_t=\beta_0+\sum_{j=1}^D \beta_j g\left(\gamma_{0j}+\sum_{i=1}^m \gamma_{ij}X_{t-i}\right)+\varepsilon_t.
\end{equation*}
A leading example for the active function $g$ is the logistic function (\ref{Logistic_function}). Figure 3 illustrates the architecture of a feedforward network with 3 input units, 4 hidden units, 1 output unit and shortcut connections.
 \def\layersep{2.5cm}
 \usetikzlibrary{calc}
\begin{figure}[ht]
\begin{center}
\begin{tikzpicture}[shorten >=1pt,->,draw=black!50, node distance=\layersep]
	\tikzset{normal arrow/.style={draw,->,>=stealth}}
    \tikzstyle{every pin edge}=[<-,shorten <=1pt]
    \tikzstyle{neuron}=[circle,fill=black!25,minimum size=17pt,inner sep=0pt]
    \tikzstyle{input neuron}=[neuron, fill=green!50];
    \tikzstyle{output neuron}=[neuron, fill=red!50];
    \tikzstyle{hidden neuron}=[neuron, fill=blue!50];
    \tikzstyle{annot} = [text width=4em, text centered]

    \foreach \name / \y in {1,...,3}
        \node[input neuron,pin={[pin edge={<-,>=stealth}]left: $X_{t-\y}$}] (I-\name) at (0,-\y) {};

    \foreach \name / \y in {1,...,4}
        \path[yshift=0.5cm]
            node[hidden neuron] (H-\name) at (\layersep,-\y cm) {};

    \node[output neuron,pin={[pin edge={->,>=stealth}]right:$X_t$}, right of=H-3] (O) {};

    \foreach \source in {1,...,3}
        \foreach \dest in {1,...,4}
            \path[normal arrow](I-\source) edge (H-\dest);

    \foreach \source in {1,...,4}
        \path[normal arrow] (H-\source) edge (O);
      
    \node[annot,above of=H-1, node distance=1cm] (hl) {Hidden layer};
    \node[annot,left of=hl] {Input layer};
    \node[annot,right of=hl] {Output layer};
    
 \draw[normal arrow,->]  
    ([yshift=0,xshift=11mm]O.north) .. controls + (0,2) and + (0,0) .. ([yshift= -2 mm, xshift=3 mm] H-1.north) node [pos=0.3,right=10pt] {Feedback error};   
\end{tikzpicture}  
\end{center}
\begin{tikzpicture}[
init/.style={draw,circle,inner sep=2pt,font=\Huge,join = by -latex},
squa/.style={
  draw,
  inner sep=2pt,
  font=\Large,
  join = by -latex
},
start chain=2,node distance=13mm
]
\node[on chain=2] 
  (x2) {$X_{t-2}$};
\node[on chain=2,join=by o-latex] 
  {$\gamma_{2j}$};
\node[on chain=2,init] (sigma) 
  {$\displaystyle\Sigma$};
\node[on chain=2,squa,label=above:{\parbox{2cm}{\centering Activate \\ function}}]   
  {$g$};
\node[on chain=2,label=above:Output,join=by -latex] (O)
  {$X_t$};
\begin{scope}[start chain=1]
\node[on chain=1] at (0,1.5cm) 
  (x1) {$X_{t-1}$};
\node[on chain=1,join=by o-latex] 
  (w1) {$\gamma_{1j}$};
\end{scope}
\begin{scope}[start chain=3]
\node[on chain=3] at (0,-1.5cm) 
  (x3) {$X_{t-3}$};
\node[on chain=3,label=below:Weights,join=by o-latex] 
  (w3) {$\gamma_{3j}$};
\end{scope}
\node[label=above:\parbox{2cm}{\centering Bias \\ $\gamma_{0j}$}] at (sigma|-w1) (b) {};

\draw[-latex] (w1) -- (sigma);
\draw[-latex] (w3) -- (sigma);
\draw[o-latex] (b) -- (sigma);

\draw[decorate,decoration={brace,mirror}] (x1.north west) -- node[left=10pt] {Inputs} (x3.south west);
\draw [-latex] (O) - ++(0,-1) |- (w3) node[pos=0.7,below] {Feedback error};

\end{tikzpicture}
\caption{A feedforward network with $m=3$ input units, $D=5$ hidden units and 1 output unit.}
\end{figure}

\section{Empirical Results}
\subsection{Nonlinear Time series Analysis}

We perform a Ter\"{a}svirta  test to detect the presence of a Logistic smooth transition model. The test is based on a Taylor series expansion of the general LSTAR model. 

We take the third order Taylor approximation of  the Logistic function (\ref{Logistic_function}) with respect to $h_t=-\gamma(t-c)$ with threshold variable $Z_t=t$ evaluated at $h_t=0$. The expansion has the form:
\begin{equation*}
G(t;\gamma,c)\simeq\frac{h_t}{4}-\frac{h_t^3}{48}
\end{equation*}
so that
\begin{equation*}
X_t=\phi+\phi_0 X_{t-1}+\ldots \phi_p X_{t-p}+ (\beta+\beta_0 X_{t-1}+\ldots \beta_p X_{t-p})(\frac{h_t}{4}-\frac{h_t^3}{48})+\varepsilon_t. 
\end{equation*}

The first step is to estimate the linear portion of the AR(p) model to determine the order p. A $p$ order of one or zero AIC and BIC respectively (see Table \ref{AIC}). 
\begin{table}[H]
\centering
\caption{\small Order selection for volatility.  Best five AIC and BIC out of the first 20 lags.}
\label{AIC}
\begin{tabular}{rrr}
  \toprule
  p & AIC & BIC \\ 
  \midrule
   0 & -1308.18 & -1810.94\\ 
   1 & -3125.22 & -1804.89\\ 
   2 & -3123.25 & -1799.36\\ 
   3 & -3121.80 & -1793.28\\ 
   4 & -3119.81 &  -1787.21\\ 
\bottomrule
\end{tabular}
\end{table}
We next select the functional form. Consider two LSTAR models with order zero and one respectively :

\begin{align*}
\text{Model 1:  }&y_t=\pi_{10}+\pi_{11}y_{t-1}+G(t)\pi_{20}+\varepsilon_t \\
\text{Model 2:  } &  y_t =\pi_{10}+\pi_{11}y_{t-1}+G(t)(\pi_{20}+\pi_{21}y_{t-1})+\varepsilon_t \\
\end{align*}

From the Taylor series expansion for a zero-order LSTAR model, we need to regress the residuals from the linear model on the regressors (i.e, a constant and $X_{t-1}$) and $t$, $t^2$ and $t^3$. The estimated auxiliary regression is:
\begin{equation*}
\varepsilon_t=\underset{(4.876\times 10^{-3})}{6.577\times 10^{-3}}+\underset{(0.077)}{0.985}X_{t-1}+\underset{(9.697\times 10^{-5})}{2.128\times 10^{-5}}\,t-\underset{(4.605\times 10^{-7})}{2.619\times 10^{-7}}\,t^2+\underset{(5.918\times10^{-10})}{4.201\times^{-10}}\,t^3
\end{equation*}
The F-statistic for the entire regression is 6794; with four numerator and 434 denominator degrees of freedom, the regression is highly significant. However, the probability value of F-statistic for the null hypothesis that $t = t^2=t^3=0$ in the auxiliary equation is 0.2547. Hence, there is weak evidence of nonlinear behavior.

From Taylor series expansion for a first-order LSTAR model, we need to regress the residuals from the linear model on the regressors (i.e, a constant and $X_{t-1}$) and $t$, $t^2$ and $t^3$ multiplied by the regressors. The estimated auxiliary regression is:
\begin{equation*}
\varepsilon_t=\underset{(0.010)}{0.004}+\underset{(0.039)}{0.952}X_{t-1}+\underset{(0.131)}{0.920}X_{t-1}\,t-\underset{(6.090\times 10^{-4})}{6.54\times 10^{-6}}X_{t-1}\,t^2+\underset{(4.169\times10^{-9})}{9.471\times 10^{-9}}X_{t-1}\,t^3
\end{equation*}
The F-statistic for the entire regression is 6973; with four numerator and 434 denominator degrees of freedom, the regression is highly significant. Moreover, the F-statistic for the presence of the nonlinear terms $X_t\,t$, $X_t\,t^2$ and $X_t\,t^3$ is 5.14; with three numerator and 434 denominator degrees of freedom, we can conclude that there is STAR behavior. Next, we can determine if LSTAR or ESTAR behavior is the most appropriate. Given that the probability of the t-statistic on the coefficient for $X_{t-1}\,t$ is 0.00412, we cannot exclude this expression from the auxiliary equation. Hence, we can rule out ESTAR behavior in favor of LSTAR behavior. 

The coefficients of the LSTAR model are estimated using non-linear least squares. The gamma parameter is estimated by means of a grid search ranging from 1 to 200 with step increament 0.002 and initial value 3.
\begin{equation*}
X_t=\underset{(0.004)}{0.007}+\underset{(0.008)}{0.990}X_{t-1}+ \underset{(0.022)}{(-0.114}X_{t-1})/(1+\exp(\underset{(205.18)}{-9.15}(t-\underset{(3.86)}{167.26}))+\varepsilon_t.
\end{equation*}
Notice that the estimated standard deviation of the gamma parameter is very large and the coefficient of the first lag in the low regime is closely to one. 

To determine whether the two regime LSTAR is the most appropriate model, we compare few non-linear models against LSTAR in terms of Akaike and Bayesian Information criteria (AIC and BIC respectively) and Mean Absolute Percentage Error (MAPE).  

A summary of the results of applying the various models to the Red Hat volatility is shown in Table \ref{models}. All models are effectively fitting the volatility in terms of MAPE.

\begin{table}[ht]
\centering
\caption{Non-linear models Red Hat stock volatility}
\scriptsize
\begin{adjustwidth}{-1cm}{}
\begin{tabular}{cccccc}
\toprule
 Model		& Intercept & First Lag $X_{t-1}$ & AIC &BIC& MAPE \\ 
\midrule
Linear & 0 & 0.9923$^{***}$& -3129 &-3121 & 3.05 \% \\
LSTAR (2 regimes)& & &-3158&-3133 & 4.38 \% \\
 SETAR (3 regimes)& & & -3199 & -3166& 4.52 \%  \\ 
 Low regime & 0 & 0.7926$^{***}$ &&\\
 Middle regime & 0.1018$^{***}$ & 0.8594$^{***}$ &&\\  
High regime & 0.0151$^{***}$& 0.8761$^{***}$&&\\ 
Threshold & Values & Prop. in Low& Prop. in Middle& Prop. in High&\\
$Z_t=time$& 85 167& 19.36\% & 18.68 \% &61.96\% \\
LSTAR (3 regimes) &&&-3151&-3110& 4.44\% \\
Low regime & 0 & 0.9904$^{***}$ &&&\\
 Middle regime &0 & -0.114$^{***}$ &&&\\  
High regime & 0.0423$^{***}$& -0.3634$^{***}$&&&\\ 
Smoothing Parameter&gamma= 24.69 59.89&&&\\
Threshold & Values & Prop. in Low& Prop. in Middle& Prop. in High&\\
$Z_t=time$&167 397& 19.36\% & 18.68 \% &61.96\% &\\
ANN &1-2-1 with 7 weights &&-3119&-3090& 2.98 \%\\
\bottomrule
\multicolumn{3}{c}{$^{***}$ Indicates significant at 0.0001\%.}\\
\end{tabular} 
\end{adjustwidth}
\label{models}
\end{table}
The three regime SETAR model is the best in terms of AIC and BIC with a value of -3199 and -3166 respectively. However, it does a bit worse than the two regime LSTAR by about 0.1\% MAPE. The neural net 1-2-1 with 7 weights faired the lowest MAPE of 2.99 \%. It however performes relatively poorly in terms of information criteria. Hence, this comparation suggests a three regime SETAR versus a two regime LSTAR.

We examine various grapical analysis. Some of the results relating to the SETAR model are shown in Figure.

\begin{figure}[H]
\begin{center}
\subfigure[t][Residuals of a three regime SETAR.]{
\resizebox*{5.9cm}{!}{\includegraphics[scale=0.15]{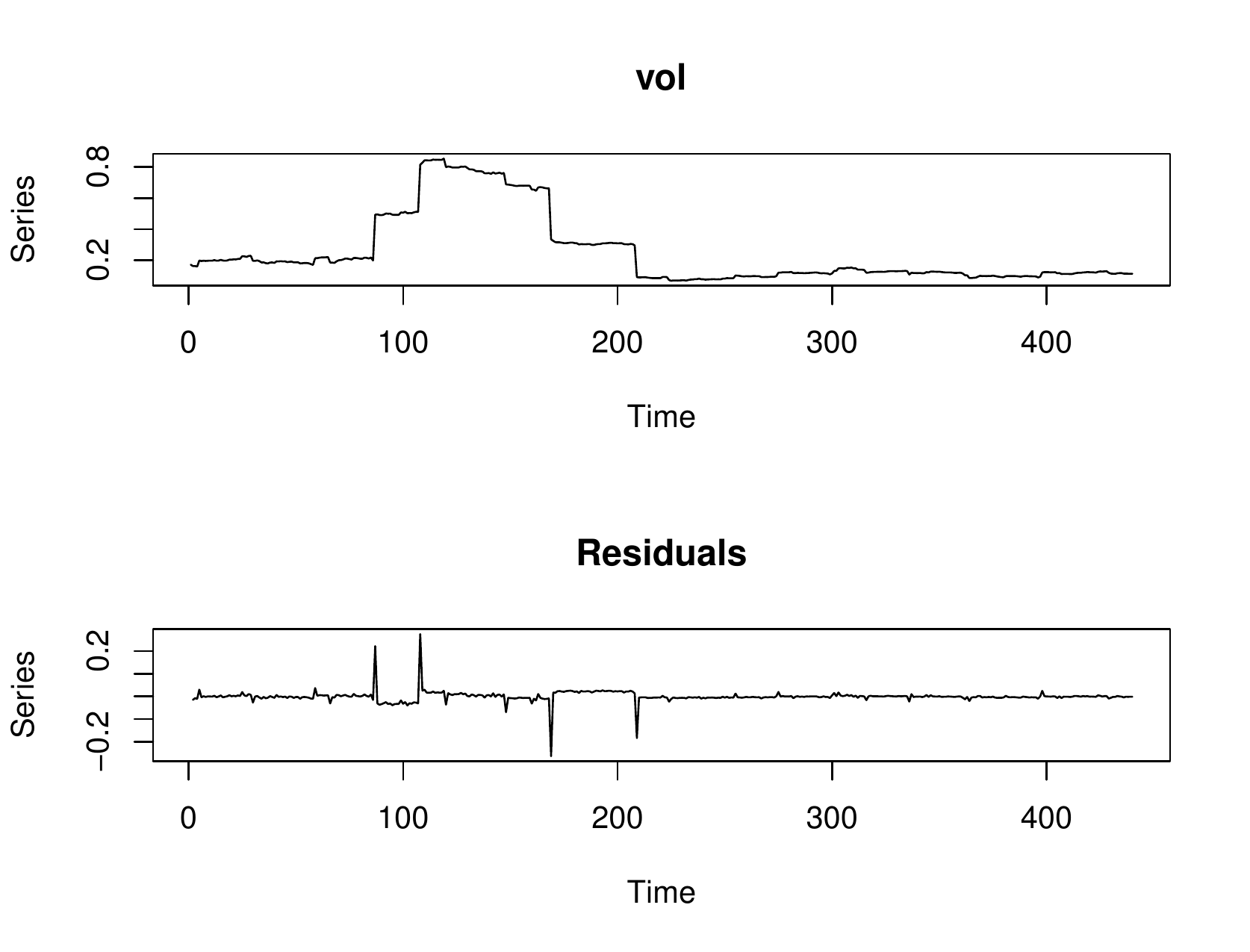}}}\hspace{-5pt}
\subfigure[t][Autocorrelations RedHat volatility.]{
\resizebox*{5.9cm}{!}{\includegraphics[scale=0.15]{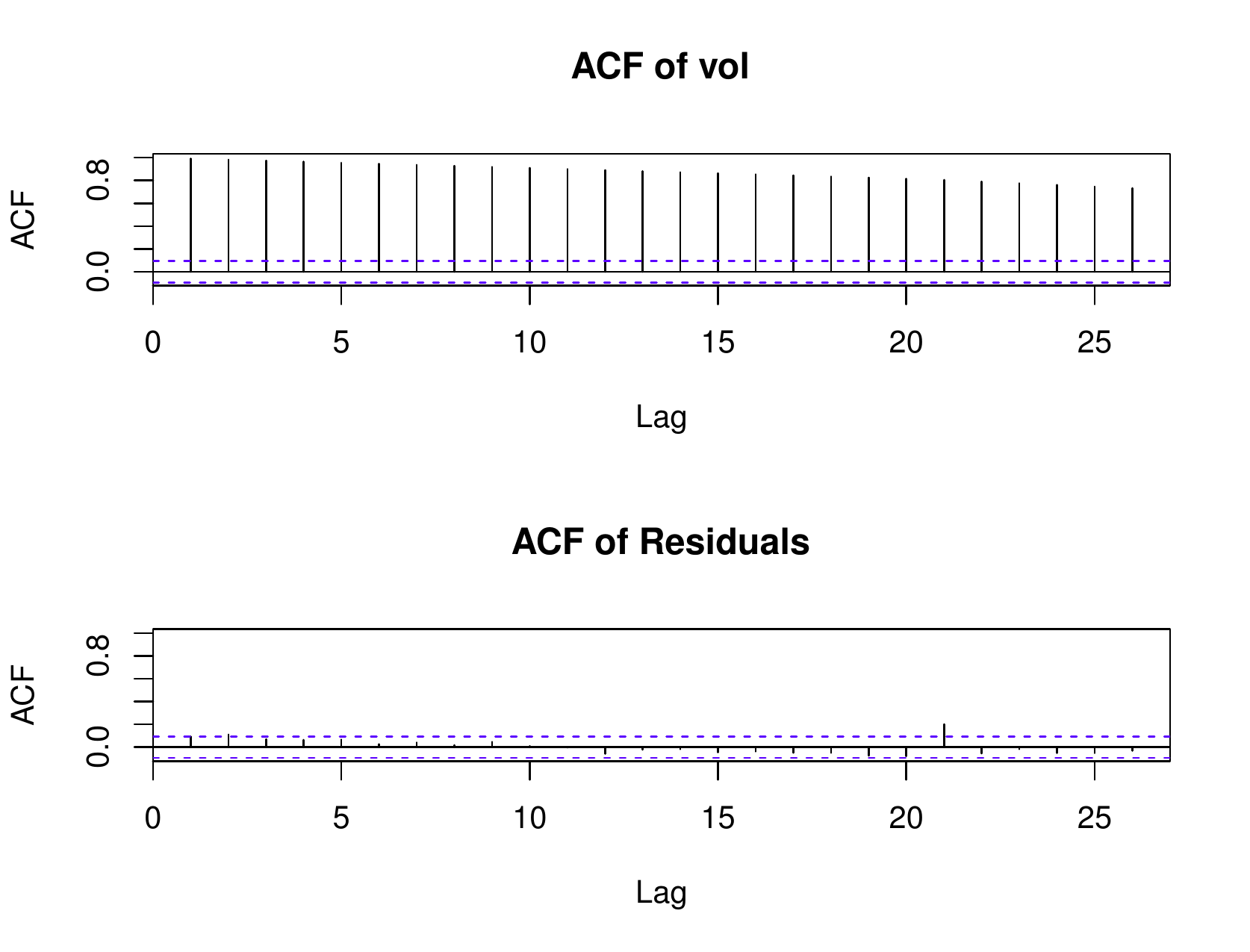}}}
\caption{}
\label{RH_series}
\end{center}
\end{figure}

\begin{figure}[H]
\begin{center}
\includegraphics[scale=0.4]{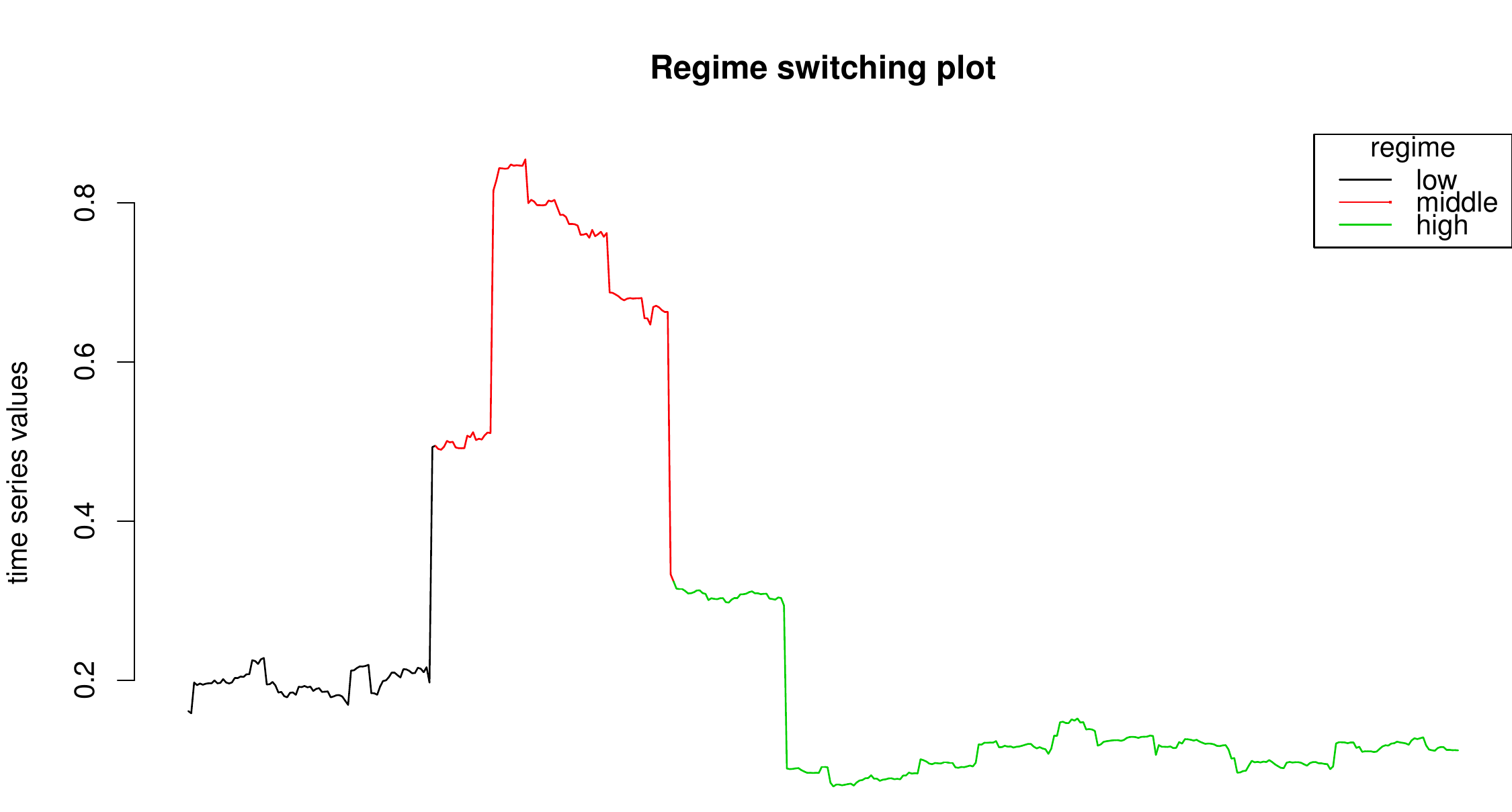}
\caption{}
\end{center}
\label{RH_Volatility}
\end{figure}


\section{Conclusions}
We examine wether there is a reduction in the Red Hat inc.  stock volatility during the moving from NASDAQ to New York Stock Exchange (NYSE) on December 12, 2006. We used a variety of non-linear time series models which included the following: self-exciting transition regression models, logistic smooth transition and artifial neural networks. The Akaike and Bayesian information and the mean absolute percentage error in forecasting were used to compare across models. All models performed pretty well in terms of MAPE with differences between 1.5\% and 0.05\%.

The self-exciting transition with three regimes model was clearly the best option in terms of AIC and BIC. The fitted model captures all the features of the data except the jump in the price of the Red Hat stock due to the announcement and change of the financial markets. This is reflected in the volatility residuals with four jumps, see Figure 4(a).


\bibliographystyle{plain}

\bibliography{PhD_books_cites} 

\begin{thebibliography}{1}

\bibitem{franses2000non}
Philip~Hans Franses and Dick Van~Dijk.
\newblock {\em Non-linear time series models in empirical finance}.
\newblock Cambridge University Press, 2000.

\bibitem{morlanes2009empirical}
Jos{\'e}~Igor Morlanes, Antti Rasila, and Tommi Sottinen.
\newblock Empirical evidence on arbitrage by changing the stock exchange.
\newblock {\em Advances and Applications in Statistics}, 12(2):223--233, 2009.

\end{thebibliography}
\end{document}